\documentclass[twocolumn,showpacs,preprintnumbers,amsmath,amssymb]{revtex4}

\usepackage{graphicx}
\usepackage{dcolumn}
\usepackage{bm}
\usepackage{footnote}

\begin{document}

\title{Andrzej P\c{e}kalski networks of scientific interests with internal degrees of freedom through self-citation analysis}

\author{M. Ausloos$^{1}$ \email{Marcel.Ausloos@ulg.ac.be}}

\author{R. Lambiotte$^{1}$ \email{Renaud.Lambiotte@ulg.ac.be}}

\author{A. Scharnhorst$^{2}$ \email{andrea.scharnhorst@vks.knaw.nl}}

\author{I. Hellsten$^{2}$ \email{iina.hellsten@vks.knaw.nl }}

\affiliation{ $^{1}$ Group of Researchers for Applications of Physics in Economy and Sociology, GRAPES, B5a Sart-Tilman, B-4000 Li\`ege, Belgium}

\affiliation{
$^{2}$ The Virtual Knowledge Studio for the Humanities and Social Sciences at the Royal Netherlands Academy of Arts and Sciences, VKS-KNAW, Cruqiusweg 31, 1019 AT Amsterdam, The Netherlands
}

\date{\today}

\begin{abstract}
Old and recent theoretical works by Andrzej P\c{e}kalski (APE)   are recalled as possible sources  of interest for describing network formation and clustering in complex (scientific) communities,  through self-organisation and percolation processes. 
Emphasis is placed on APE  self-citation network  over four decades.
The method is that used for detecting scientists' field mobility by focusing on author's self-citation, co-authorships and article topics networks as in \cite{hellsten1,hellsten2}.  It is shown that APE's self-citation patterns reveal important information on APE interest $for$ research topics over time as well as APE engagement $on$ different scientific topics and $in$ different networks of collaboration.  Its interesting complexity results from ''degrees of freedom'' and external fields leading to so called internal shock resistance.  It is found that  APE network of scientific interests belongs to independent clusters and occurs through rare or drastic events as in irreversible ''preferential attachment processes'', similar to those found in usual mechanics and thermodynamics phase transitions.  
\end{abstract}

\maketitle

\section{Introduction}

Self-citation analysis is part of wide bibliometric analysis of scientific and scholarly citation patterns  \cite{hellsten1,hellsten2}. Often in much of the recent literature in citation analysis \cite{wouters1,h}, author's self-citations are excluded as 'noise' or  are treated as a bias for the analysis (e.g. \cite{loet1,macroberts1,persson1}), whence contempted or used to draw a negative conclusion on an author activity.    We disagree with such a line of thought. As an example, consider Andrzej P\c{e}kalski self-citations in his published work over his present career.

Andrzej P\c{e}kalski was born in Warsaw  on Nov. 02, 1937. After $the$ war he moved with his parents to Wroclaw. He graduated from the University of Wroclaw in 1960, and got a Ph.D. in Theoretical Physics from the Academy of Sciences Low Temperature and Structural Research Institute, in Wroclaw, in 1970. One of us (MA) met Andrzej P\c{e}kalski (APE) at a MECO conference in
Bled (then in Yugoslavia), in 1976. Both were interested in {\it Random Spin Systems} (RSS) and quickly published a paper on Physical Properties of a Spin Model described by an Effective Hamiltonian with Two Kinds of Random Magnetic Bonds  \cite{1,MAAPE160}, Fig.1. APE introduced MA to his wife, Joanna, and to one of his friends Jacek M. Kowalski, among others. MA also introduced his wife in their scientific activities. Within a close   family-like cluster they published several papers resulting from the study of the so called {\it Magnetic Lattice Gas} (MLG) \cite{MLG}, i.e. inserting on the classical lattice gas model a new degree of freedom, a spin, for the basic entity. This leads to the study of the Blume-Emery-Griffiths model, with particular interactions, in view of obtaining some information on the phase diagram of ferrofluids \cite{44,66,67}, and  other rather unusual disordered systems \cite{45}, often within mean field approximations \cite{MLG,MDAPERNG}.

\begin{figure}
\caption{\label{fig1}    Marcel Ausloos and Andrzej P\c{e}kalski seen discussing  Physical Properties of a Spin Model or Oxygen Diffusion in YBCO, at a Karpacz winter school   \cite{MAAPE160} }
\end{figure}

Going away from the study of static properties, of a fluid, seems a logical step, followed by APE, who started next to work on two-dimensional Diffusion (2DD) \cite{bookDiffusion}.  E.g., with one of the authors (MA),  he studied extensively the tracer (surface) diffusion coefficient of oxygen in  deficient CuO planes of YBCO \cite{MAAPE160,AP94,AP95}. Most likely this led him to study molecular \cite{21} phase formation (and reactive)  trapping, and by extension, entity trapping, thus predator-prey models \cite{drozAPE,APCM04} 
and population evolutions or dynamics \cite{APE98,MPSWPRL76,MMPVgrevol,KSWAPE99}. Such biophysics considerations are being pretty close to or entangled into its macroscopic counterpart, the behavior of human population \cite{KSWAPE99}, APE went on to study {\it Bio-Socio-Econo} (BSE)-$physics$  problems, like Model  of Wealth and Goods Dynamics in a Closed Market \cite{99} and recently prison riots \cite{riots}.

APE scientific journey during the last four decades, both in terms of number of publications or collaborations and moves to different research fields, can be more adequately followed by  focussing on how the network of citations to his own published papers develops over time. These self-citations may carry important information on how the scientist sees his own work within his/her  investigations at the time of self-citation. We do not answer the question whether self-citation $pays$ or distorts the reliability of scientific impact measures \cite{fowler}. We will show that we disagree with suggestions to remove self-citations \cite{fowler}  from citation counts. On the contrary we confirm that self-citations can be used for understanding the network structure of an author \cite{somasanyal}.  We have already shown  \cite{hellsten1,hellsten2} that there is some interest  in examining self-citations in order to provide some insight on the creativity and change of field of interest in a scientist. To do so, we use a recently developed method to analyze self-citations in combination with co-authorships (and keywords) in the self-citing articles as a potential tool for tracing scientists' field mobility \cite{hellsten1,hellsten2} based on a de-clustering method finding its root in percolation theory ideas \cite{lambi1,lambi2}. Moreover this sort of consideration ties the present research to the newly expanding research field about  social communities \cite{newmann}, in which questions pertain to the network number of links, triangles, ... and the modelization of complex (social or other) networks in order to represent statistical features, evolution, ... - not mentioning the questions of universality and robustness.

Subsequently, our research questions are: 
\begin{enumerate} \item  Which structures do APE self-citation networks entail? 
\item  Can these structures be interpreted in terms of research topics or fields?
\item Can self-citation networks be used as a means to uncover  APE field mobility, if any?
\item Are changes in  APE co-authorship associated with changes in the self-citation networks?  
\item Do emerging scientific collaborations in terms of possible changes in the co-authorships also indicate something about APE focus of study?
\end{enumerate}

The publication record of APE is of interest for doing so, in view of recent studies on other authors, like Werner Ebeling \cite{hellsten1,hellsten2}, for three reasons: First, both have founded or led a scientific school in theoretical physics. Second, it is ''known'' (see above for APE)  that both have been engaged in networks of changing collaborators over time, and, third, their  publication record is markedly different.  In this paper we take self-citations   as a source of  information on the development of the scientific career of APE, quantify and visualize his scientific journey through  research fields. There will be no consideration on some ''work quality'', the definition being unclear to the authors, nor about the number of citations by others, nor any {\it impact factor}.

The paper is organized as follows. In section II, we introduce the data set and the method used.  In section III, we present our results obtained using the method. Thereafter, in section IV, we discuss the results.

\section{Data and Method}

In a recent CV, released in April 2007, APE claims to have 90 papers in (we quote him) ''international reviewed journals''. In fact, his publication record can be downloaded in several ways : first from his webpage \cite{APEwebpage}, second from the {\em Web of Science}, (the {\em Science Citation Index}, the {\em Social Sciences Citation Index} and the {\em Arts and Humanities Index}), using the Boolean search operation  {\it Pekalski A}   (not P\c{e}kalski ) in the author field (General search interface). The APE web page indicates  90 articles since his first in 1966 \cite{APE1}. The {\em Web of Science} (accessed, May 08, 2007) surprisingly recalls 95 such articles since 1969 \cite{APEwebscie}. This record, of course, only encompasses articles that are indexed in the citation index databases, i.e.  we  excluded from this analysis : (i) articles published in  not ISI-indexed \cite{ISI} journals, (ii)  not yet published papers at the time of publication of  the latter, (iii)   books and book chapters. 
Self-citations here are taken as the  papers with APE as one of 
the authors which cite other papers with APE as one of the authors.
We do not normalize the number of self-citations, neither  e.g. to the number of papers of the author of coauthors, nor to the number of citations in one or the whole paper(s). Let us note that ISI has also backwards indexed all authors, so one gets all papers of Pekalski, no matter if he stands first or second or .... author.

There are several possible sources of disagreement between the data ''banks''
\begin{enumerate} \item  papers are not in the Web of Science because they did not appear in ISI journals.
In this case the work can only be improved if for these papers the reference lists are available.
 
 \item  papers from the list which are in ISI publications have not been found because of misspelling or different spelling of the name of the author.  P\c{e}kalski  might sometimes appear as $P_{-} kalski$ , but  $Pekalski$ is not so a troublesome case,- we have only found a homonym
A. P\c{e}kalski  in Delft, in engineering. In fact, such possible
 false records usually get sorted out by not being linked to the
self-citation network of the main target. Since all co-authors are indexed,   one will not likely miss papers because "Pekalski" is not the first author.
\end{enumerate} 

In fact the authors are aware that APE webpage also misses some articles, like \cite{MAAPE160,SCAPElncs04}, the second one \cite{SCAPElncs04}  being surely  in the ISI citation index \cite{ISI} the impact factor of the ''journal'' in which it is published being e.g.  0.513 in 2004 \cite{LNPisistop90}. 
 
 Notice that we have $not$ searched for nor considered papers published in books or proceedings, articles which could be also self-quoted. {\it A posteriori}, it  seems that such a slight neglect has not influenced the conclusions to come.
Therefore in the following we use the 95 {\em Web of Science} articles as the empirical data basis.

In this article, we use a specific method   that focuses on percolated islands of nodes, is called the Optimal Percolation Method (OPM) \cite{hellsten1,hellsten2}, and is a variant of percolation idea-based methods (PIBM) used in order to reveal structures in complex networks \cite{lambi1,lambi2}.  One of the advantages of OPM is the rapid identification of the resulting division of the whole network into sub-structures.  

Recall that the  OPM differs from the PIBM in at least two ways.
 On one hand, OPM applies to unweighted networks, while PIBM apply to correlations matrices (=weighted networks). On the other hand, in OPM, one removes nodes instead of links. 

On OPM, we define the connectivity $c$ of the network as the number of pairs of nodes that can be joined by a path (of arbitrary length).  By definition, $c$ is equal to $N  (N-1)/2$, where $N$ is the number of nodes in the connected network. Then, we search for $k$ nodes so that when they are removed from the network, the connectivity $c$ is minimal. These $k$ nodes define the intersection of structures that we identify as different sub-groups. Finally, we plot the uncovered sub-groups in different   tones for the sake of clarity. 

There is therefore also a huge difference between OPM and PIBM. In PIBM, one removes links depending on their value and then observes how the system breaks into clusters. In OPM, in contrast, one looks for the nodes that, if removed, optimize the breaking of the system, and one removes them. 

We concentrate here below on the appearance of 'clusters', that means groups of articles which are more intensively linked to each other than to the rest. We interpret this clustering as a sign that the articles in a sub-group have something in common. 

The analysis proceeds in five steps: First, we take a look at the frequency of the  ISI-indexed articles authored or co-authored by APE over time. Second, the overall structure of his self-citation network is examined. Third, we apply the OPM method to this network. Next, we analyze the co-authorships in these articles. Finally, we show the overall development over article types as a function of the year of publication and the detected topics.   

\section{Results}

\subsection{Basic Statistics and Clusters }

We start with the empirical time evolution of the number of published articles by APE (and his co-authors) as a function of the year of publishing the article (Fig. 2). We see that APE has been continuously productive during the last 40 years. His yearly productivity fluctuates between 0 and 9 articles, with an average around 3 papers/year.  He has sometimes gaps in his production, but his productivity takes off in 1980 and accelerates at the end of last century (Fig. 3).  Observe the plateau ending in 1994 or so.

The total number  of APE co-authors  is equal to 11;  the number  (20) of single-authored papers stays remarkably persistent over time, - in this list. The network of his co-authors varies and of course grows over time. Whence to further reveal the structure of APE scientific journey  it is of interest to look for the $position$  of the articles in his self-citation network. 

\begin{figure}
\includegraphics[angle=-90,width=3.3in]{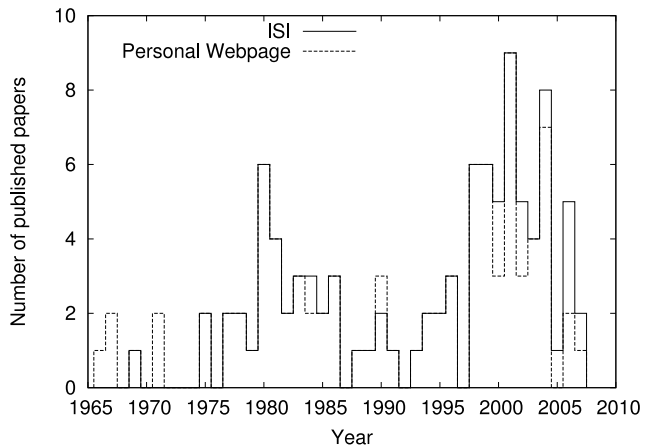}
\caption{\label{fig2}  Number of published papers in a given year  written by 
Andrzej P\c{e}kalski:  black line: (95) ISI indexed articles; dashed line: 
(90) from his web page  }
\end{figure}

\begin{figure}
\includegraphics[angle=-90,width=3.3in]{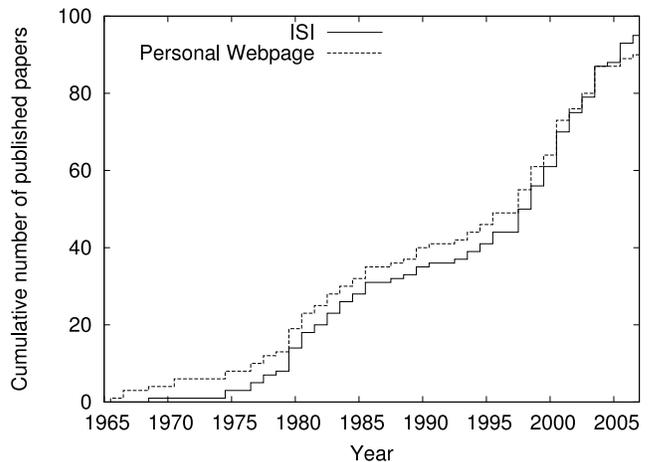}
\caption{\label{fig3}  Cumulative distribution of the 95 ISI indexed articles written by Andrzej P\c{e}kalski 
as a function of time; dashed line: same for the 90 articles mentioned on his web page}
\end{figure}

As a next step in our analysis, we apply the Optimum Percolation Method algorithm to the percolated island\footnote {In the case of Ebeling \cite{hellsten1,hellsten2} it was necessary to consider values of the network node degree $k$ smaller or equal to 4 in order to highlight the clusters; here $k$=0}. We deduce 3 sub-groups, as verified in Fig. 4. The structures, plotted in grey,black, and white, are composed of 9, 7 and 32 nodes, respectively, and 18, 19 and 68 links (citations {\it from this list}) respectively. Note, that the clusters represented in Figure 4 should not be read as evolutionary trees. Different subgroups contain articles from different (years) points in time. Later on, we will make also the temporal structure of the self-citation network visible. 

For completeness, let us observe what papers are most often quoted, i.e. what seems the most relevant ones for the authors; those who get the most links, and how many
\begin{enumerate}

 \item 6 times:   A. Pekalski and M. Ausloos:  Physical Properties of a Spin Model described by an Effective Hamiltonian with Two Kinds of Random Magnetic Bonds \cite{1}
    
     \item 
  5 times:  M. Ausloos, P. Clippe, J.M. Kowalski, and A. P\c{e}kalski: Magnetic Lattice Gas \cite{MLG}

\item 17 times:  I. Mroz, A. Pekalski, and K. Sznajd-Weron: Conditions for adaptation of an evolving population  \cite{MPSWPRL76}
\end{enumerate} 

In order to characterize the trends  of activity or research fields in the career of APE, as well as to verify the pertinence of the automatic classification, we can take a look at the keywords associated to the articles in each cluster, when available. There are only a few overlaps between the three clusters;  this confirms the relevance of the three revealed sub-groups. Therefore clusters in the self-citation network(s) can be used to demark different research fields, whence the creativity and adaptation of the author.  

The interpretation of the three different structures is not difficult if one is acquainted with APE work. The first (grey plotted) area is related to articles written about {\em random/disordered spin models}.  Work in this area belongs to classical  streams in statistical mechanics theory. The second area (black plotted) strictly contains  work on the MLG physics. This belongs to models in which an extra degree of freedom allows coupling with new external fields and widely extends the phase diagram \cite{MEM,MDLA}.  Interestingly one could imagine a strong connexion with the RSS model works; it is not for P\c{e}kalski. The third  and largest  area (white plotted) represent a huge branch of investigations in APE work, namely {\em diffusion, adaptation, self-organisation} research,  which is entangled in a competition of entities problems, like the prey-predator problems. These are much less specific and classic topics within statistical physics, though exist since studies in population studies and dynamics, going back to Malthus and Verhulst. It is remarkable that for P\c{e}kalski  these are connected questions. 

	Another possible difference between the identified subfields is the respective co-authorship networks in the clusters. We have identified the most present co-authors in each structure (Table \ref{tab2} ). The empirical results show that the respective lists of co-authors are  'loosely similar', namely there is no strong dependence between the research field or topics of the paper and the co-author with whom the paper was written, $-$ as deduced from this self-citation list. Nevertheless note that only a few APE co-authors  occur in three clusters, and very few in two clusters.
	
	\begin{table}[t]
\begin{tabular}{|l|c|l|c|l|c|}
\hline
 & No  &   & No  &  & No       \\
\hline
\hline
Grey  &  9  &  Black &    7  & White &  32       \\
\hline
\hline
RSS  &   (5) & MLG &   (5) & BSE & (15)       \\
\hline
\scriptsize{AUSLOOS} &	\scriptsize{4}& \scriptsize{KOWALSKI } 	&\scriptsize{4}	&\scriptsize{AUSLOOS} &\scriptsize{8} \\ 
\scriptsize{KOMODA}	&\scriptsize{2}	&\scriptsize{AUSLOOS }	&\scriptsize{3}	&\scriptsize{SZNAJD-WERON} 	&\scriptsize{7} \\
\scriptsize{ } &	\scriptsize{ }	&\scriptsize{CLIPPE }	&\scriptsize{3}	&\scriptsize{CLIPPE} 	&\scriptsize{4} \\
\scriptsize{  }	&\scriptsize{ }	&\scriptsize{DUDEK} &\scriptsize{2}&	\scriptsize{CEBRAT } 	&\scriptsize{3} \\
\scriptsize{  }	&\scriptsize{ }	&\scriptsize{PEKALSKA } &\scriptsize{2 }&	\scriptsize{MROZ } 	&\scriptsize{3} \\
\scriptsize{  }	&\scriptsize{ }	&\scriptsize{ } &\scriptsize{ }&	\scriptsize{DROZ } 	&\scriptsize{2} \\

\scriptsize{  }	&\scriptsize{ }	&\scriptsize{ } &\scriptsize{ }&	\scriptsize{SKWAREK } 	&\scriptsize{2} \\

\hline 
\end{tabular}
\caption{Most frequent co-authors appearing in the cluster structures. The column No shows the number of co-authored $articles$ by the given co-author, and  in parentheses the total number of  $different$ co-authors in the specific cluster. We only indicate those appearing at least twice in the given cluster }
\label{tab2}
\end{table}

	The list of collaborators makes the structure of the APE degrees of freedom visible. Many of these co-authors are either local students or senior colleagues; two co-authors are from the Li\`ege group. This shows the intense connexions of APE, his restricted degrees of freedom,  for this sort of work and its/his evolution.

	\subsection{Time dependence }
	
	So far, we have shown the correlation between self-citation clusters  and co-authorships in the self-citing articles of APE. In this subsection, we analyse the temporal structure, the scientific journey, in APE  research activities over time as perceived through his self-citations. As shown above, the percolation method has led to a decomposition of the self-citation network into three disjoint structures, that we represent with black, white and grey for reasons of visualization. In order to evaluate the time evolution of the author's career, we draw (Fig. 5) a series of boxes, each representing one article, from the first published paper to the last published paper, - in this list of self-citating papers. This leads to a rapid visualization of the periods of activities of the author in each subfield. We see that APE activities in different research fields are very concentrated at different periods in time. There is barely any overlap, reminding us of phase transitions; see for a related discussion on works by $many$ authors \cite{drastic}.
	
	During the 1970-80s, for example, his research was clearly directed toward the RSS, i.e. classical roads toward joining the club of fundamentalists in discrete models in statistical mechanics. Between 1985 and 1995 or so, APE was involved with quite a limited number of co-authors for these papers on MLG. In this period APE contributed to tying mathematics with physics. The spreading of the ideas around self-organization, irreversible processes and non-linear dynamics in physics, due to exo- and/or endogenous effects on whatever population of entities came after the 1984 plateau (Fig. 3). In addition to the articles analysed here, notice that APE organised several winter schools and edited several books, like \cite{MAAPE160b}.   These  new topics related to questions of biophysics, economy and social dynamics are the seats of an intense activity nowadays; the data indicates that APE can stick to a competitive field and participate in its evolution.

It is interesting to note that the transition from one subfield to the other is rather sharp and irreversible, i.e. the author  does not return to a subfield after an inactivity time, and seems to remain active in a subfield over long time periods. He resists external shocks, or maintains a ''high internal resistance'' to ''external fields''.
	
	It is also worth noting (Fig. 3) that productivity, if measured in terms of publications per year, is increasing over time: While it took APE ten years to publish the first ten ISI-indexed self-citing articles, it only took  six years to publish  ISI-indexed eighteen articles on BSE, or about 12 years for the last 24 papers. This result is consistent with a recent study showing that scientists' productivity over time increases during their career \cite{publishperish}, - up to the declining years !  
	
		\section{Discussion}

In much of the recent literature in citation analysis \cite{wouters1}, author's self-citations are excluded as 'noise' or are treated as a bias for the analysis (e.g. \cite{loet1,macroberts1,persson1}).   We use a recently developed method to analyze self-citations in combination with co-authorships in the self-citing articles of a well known author (APE) as a potential tool for tracing scientists field mobility \cite{hellsten1,hellsten2,lambi1,lambi2} and  if possible for the causes.

In the case of APE, a great number  (48), yet a small majority, of the 95 articles, in such a list,  is linked to each other by self-citation; three clusters emerge with in chronological order 9, 7 and 32 nodes or  articles  for a total of 48 articles; thus  47 are  not connected. The clusters are well separated.    Because we are interested in the network of self-citations, the disconnected articles have been excluded from the study, whence network structure. As compared to Ebeling, APE has more articles outside ''his percolated island". It is known to one of the authors (MA) that APE  is not prone to have a long list of references in his papers, whatever that may mean.

Five steps in our analysis build upon and support each other, hence result in an emphasis of consistent patterns in the development of the career of APE. 
 It is shown that APE's self-citation patterns reveal important information on APE interest $for$ research topics over time as well as APE engagement $in$ different networks of collaboration.   It is found that  APE network of scientific interests belongs to independent clusters and occurs through rare or drastic events which results from ''preferential attachment processes'', to some coauthor group, as in usual mechanics and thermodynamics formalism of phase transitions. 
In some sense it can be  conjectured that this interesting complexity results from ''degrees of freedom'' coupled to external fields leading to internal motivation, though submitted, most likely, to shock resistance. 

Indeed there is a strong connection (seen in the quite finite size of the clusters, and list of co-authors) between  the co-authorships and the topics used in the self-citing articles of the author. Altogether, these results seem to justify the use of self-citation networks as a key signature of a scientist career. In the case of Ebeling \cite{hellsten1,hellsten2}, it was found that the OPM analysis   suggests that changing co-authorships drive the changing research interests and move to new research topics. The same is true here, but in the opposite case, i.e. when there is not much change in the coauthors/groups. The Optimal Percolation Method, therefore, can well serve as a relevant tool for detecting the development of trends in a scientific community or in the scientific career of an author, highly productive or not.

It seems that in the case of P\c{e}kalski , the OPM has some statistical limitations since almost half of his papers lack self-citations whence are excluded form the analysis! It might be interesting to take a look at the topics of these not connected nodes in the network. It seems that the OPM works best in cases where the author heavily cites his/her other articles, or rather when the value of $k$ to be taken into account for breaking apart the main cluster is finite.

Moreover if  one aim is to trace scientists' careers using one address, in the search, might   exclude several articles  from the ISI data set. This is not the case to our knowledge for P\c{e}kalski, for which no address was inserted in the data extraction, nor for Ebeling, in fact, for which one address was used.

For biographical research concentrated on single authors, the method reveals interesting additional information. To interpret the motivations for the occurrence of certain research fields one has to look into the biography of an author, or have personal knowledge about the author.  External changes as political conditions, or geographical moves \cite{geogrbias}, but also visits of conferences, invited guest positions and longer stays abroad, should, or might, trigger new collaborations and new research topics which, in turn, should become visible in the patterns of self-citations. That seems to be somewhat the case here, but research on other authors should be worthwhile to confirm such effects.

{\bf Acknowledgements}
This paper is an outcome of the Critical Events in Evolving Networks (CREEN) project, funded by the EU under its 6th Framework, NEST-2003-Path-1, 012684.  Support from Actions de Recherche Concert\'ee Program of
the University of Li\`ege (ARC 02/07-293) is also gratefully acknowledged.
This paper is dedicated to Andrzej P\c{e}kalski for his 70th birthday; MA thanks APE  for his comments and friendship.

\newpage
\begin{figure}
\includegraphics[width=6.0in]{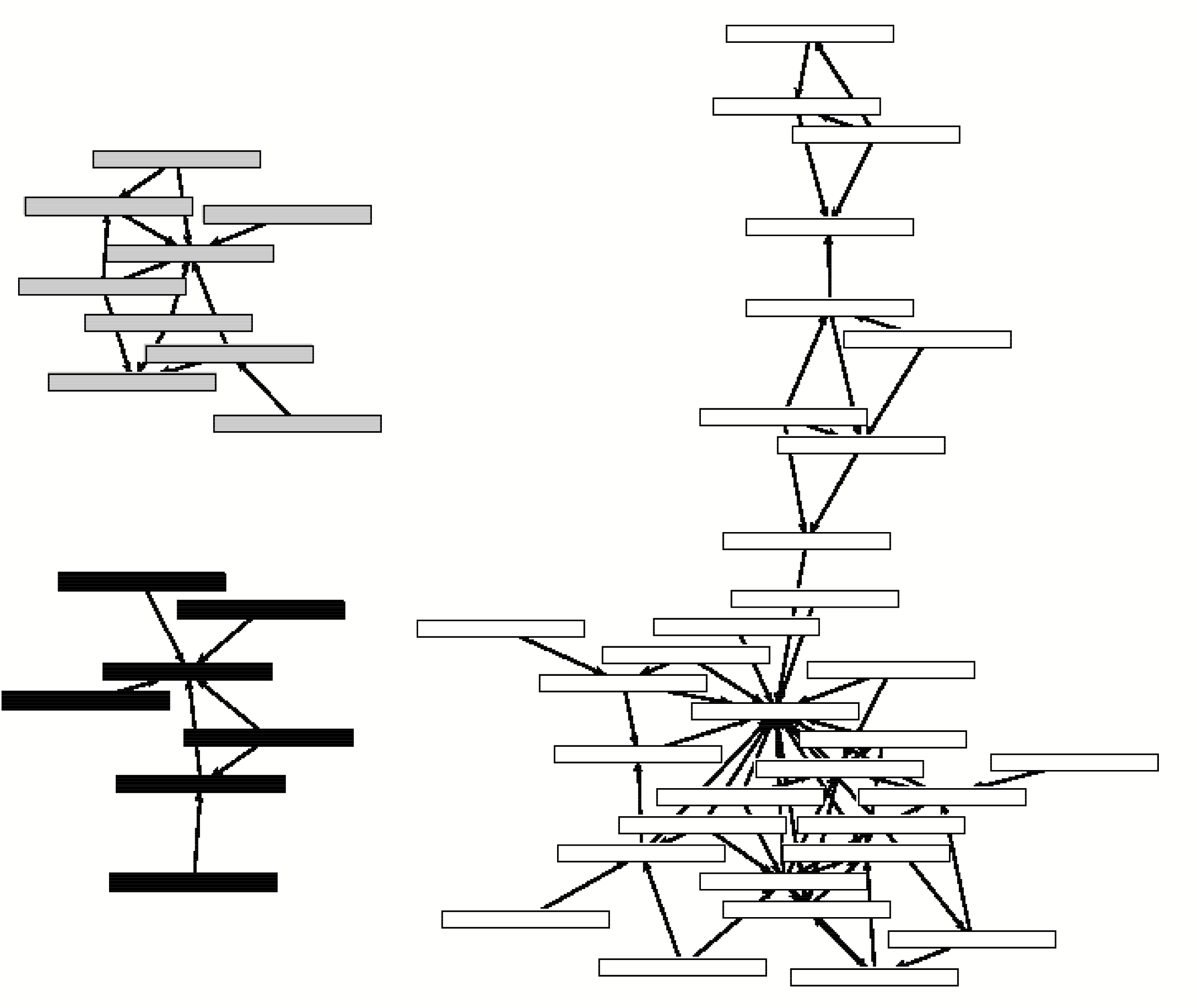}
\caption{\label{fig4} The Optimum Percolation Method applied to Andrzej P\c{e}kalski's 
self-citation network. The three revealed structures are plotted in grey 
(top left), black (lower left corner) and white (right hand side), corresponding
to Random Spin Systems (RSS), Magnetic Lattice Gas (MLG), and Bio-Socio-Econo
 (BSE)-physics  respectively }
\end{figure}

\begin{figure}
\includegraphics[width=3.3in]{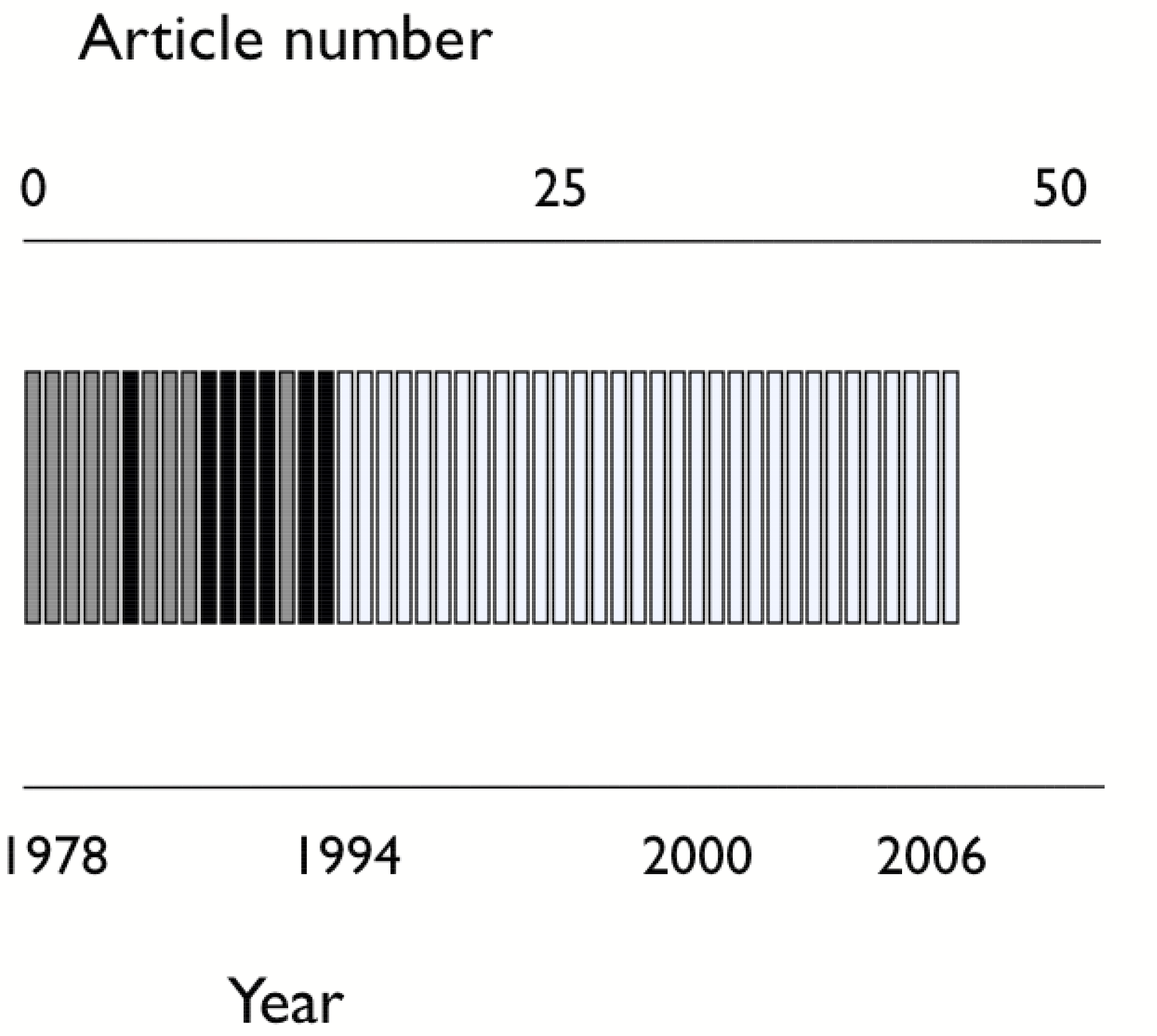}
\caption{\label{fig5} Time evolution of the article type as a function of the year/article number}
\end{figure}
\end{document}